\documentclass[structabstract,printer]{aa}
\usepackage{graphicx}
\usepackage{txfonts}
\begin{document} 
\newcommand{\be}{\begin{eqnarray}}
\newcommand{\ee}{\end{eqnarray}}
\newcommand{\vdag}{(v)^\dagger}
\newcommand{\myemail}{pfalzner@ph1.uni-koeln.de}
\newcommand {\nbody}{\textsc{\mbox{nbody6}}}
\newcommand {\nbodypp}{\textsc{\mbox{nbody6\raise.4ex\hbox{\tiny++}}}}
\newcommand {\COri} {\mbox{$\theta^1{\rm{C}}\:{\rm{Ori}}$}}
\newcommand {\rclose} {$r_{\rm{close}}$ }
\newcommand {\vclose} {$v_{\rm{close}}$ }
\newcommand {\eclose} {$e_{\rm{close}}$ }
\newcommand {\Msun} {\mbox{M$_{\odot}$}}
\newcommand {\Rsun} {\mbox{R$_{\odot}$}}
\newcommand {\Lsun} {\mbox{L$_{\odot}$} }
\newcommand {\Macc}  {$\stackrel{.}{M}_{acc}$ }
\newcommand {\Mjup} {\mbox{M$_{Jup}$}}

\title{Did a stellar fly-by shape the planetary system around Pr 0211 in the cluster M 44?}
\titlerunning{Did a stellar fly-by shape the planetary system around Pr 0211?}

\author{Susanne Pfalzner  
              \inst{1,2} \and Asmita Bhandare \inst{3} \and Kirsten Vincke  \inst{1}}
\institute{Max-Planck-Institut f\"ur Radioastronomie, Auf dem H\"ugel 69, 53121 Bonn, Germany\\
              \email{spfalzner@mpifr.de}
              \and Fresenius University of Applied Sciences, Platz der Ideen 1, D\"usseldorf, Germany
              \and Max-Planck-Institut f\"ur Astronomie, K\"onigstuhl 17, 69117 Heidelberg, Germany}

\date{}

\abstract
{}
{Out of the $\sim$ 3000 exoplanets detected so far, only fourteen planets are members of open clusters: among them an exoplanet system around  Pr 0211 in the cluster M44 which consists of at least two planets with the outer planet moving on a highly eccentric orbit at 5.5 AU. One hypothesis is that a close fly-by of a neighbouring star was responsible for the eccentric orbit. 
We test this hypothesis.}
{First we determine the type of fly-by that would lead to the observed parameters and then use this result to determine
the history of such fly-bys in simulations of the early dynamics in an M44-like environment.}
{We find that although very close fly-bys are required to obtain the observed properties of Pr 0211c, such fly-bys are relatively 
common due to the high stellar density and longevity of the cluster. About 10\% of stars actually experience a fly-by that would lead to such a
small system-size as observed for Pr0211 or even smaller. Such close fly-bys are most frequent during the first 1-2 Myr after cluster formation, corresponding to a cluster age $\leq$ 3 Myr. It is unclear whether planets generally form on such short timescales. However, afterwards the close fly-by rate is still 0.2-0.5 Myr$^{-1}$, which means extrapolating this to the age of M44  12\%-20\% of stars would experience such close fly-bys over this timespan.}
{Our simulations show that the fly-by scenario is a realistic option for the formation of eccentricity orbits of the planets in M44. The occurance of such events is relatively high leading to the expectation that  similar systems are likely common in open clusters in general.  }


\maketitle
\section{Introduction}
Ever since the first exoplanet was detected, the question arose whether planetary systems are equally common around cluster stars as those orbiting field stars.  Although several planets have been detected,  it is still far from clear whether planets are as common around stars in clusters as around field stars or not. Out of the $\sim$3000 exoplanets detected so far, only fourteen have been discovered in open clusters and  the number of detected planets varies considerably from cluster to cluster: For example,  three planets were detected in Hyades \citep{Sato:2007,Lovis:07,Quinn:2014}, five in M 44 - Praeceps \citep{Quinn:12,malavolta:16,obermeier:16}, and four in  M 67 \citep{Brucalassi:2017}, whereas, for example, no planets were detected so far in 47 Tucanae \citep{Gilliland:00,weldrake:05}, NGC 2301  \citep{howell:05} and NGC 7789  \citep{bramich:06} and other investigated clusters \citep{Nascimbeni:12}.  Therefore these results are interpreted by the respective authors from "planets are very rare" in clusters  to "they are as common or even more than around field stars". In addition, the comparatively large radii of some of the  planets in Praecepes, Hyades and Upper Scorpius indicate systematic differences in their evolutionary states or formation \citep{obermeier:16}.

The reasons why planets could be less common in open clusters than around field stars are that
\begin{itemize}
\item disc destruction during the first few Myr could prevent planet formation,
\item already formed planetary systems could be destroyed in the dense duster environments in the consecutive Gyrs.
\end{itemize}

\begin{table*}[t]
% \begin{indented}
\begin{tabular}{@{}lllllll}
\hline
 Parameter & P [days]& $a$(AU) &  e & $\omega$ (deg) & $M_p \sin i$(\Mjup) \\
\hline
  Pr0211b    & 2.14610 $\pm$ 3 $\cdot$ 10$^{-5}$ & 0.03176 $\pm$ 0.00015 & 0.011$^{+0.012}_ {-0.008}$ & 17$^{+87}_ {-111}$ &
 1.88 $\pm$0.03 \\
Pr0211c    & 4850$^{+4560}_ {-1750}$              &  5.5$^{+3.0}_ {-1.4}$ & 0.71$\pm$0.11                         &111$\pm$9 & 7.79$\pm$0.33 \\
\hline
\end{tabular}
%\end{indented}
 \caption{\label{Table.One}Orbital parameters of the two planets orbiting Pr0211 in M44 \citep{malavolta:16}}
\end{table*}

Disc destruction can happen either by stellar fly-bys  or external photo-evaporation by nearby massive stars. An extensive body of theoretical studies of the influence of both effects in clusters of various densities exists \citep{alexander:06,ercolano:08,drake:09,gorti:09,adams:10,Dukes:12,steinhausen:14,adams:15,vincke:16}. It seems that the influence on the frequency of discs  and the planetary systems is moderate in relatively short-lived clusters or associations,  like NGC 2024, IC348 or even the ONC, typical for the solar neighbourhood \citep{malmberg:09,malmberg:11,simon:15}. Nevertheless the actual properties of the disc and  the resultant planetary system might be strongly influenced in these environments\citep{vincke:16}. The situation is less clear for long-lived open clusters,  like M44. Obviously  once these clusters were more compact so that they were able to survive the violent early phase of their development. Young counterparts ($<$ 5 Myr) of long-lived open clusters are, for example, Arches (Stolte et al. 2010), NGC 3603 \citep{roman:16} and or Trumpler 14 \citep{mesa:16}. Recent simulations have tried to determine the fraction of planets that become affected by the cluster environment and either move on an eccentric orbit or become unbound \citep{Hao:2013,Li:2015}. The necessarily very high stellar density means the influence of the environment is much stronger and disc destruction more likely than for the short-lived clusters \citep{vincke:15}.

Here we want to concentrate on the exoplanets found in the open
clusters M44 (also referred to as Praesepe or NGC 2632),  in particular on the two planets orbiting the star Pr 0211, which is the only planetary system found in a cluster so far. 
The cluster M44 is located at a distance of 187 pc and has an age of 790 $\pm$ 30 Myr estimated by isochrone fitting (Brandt \& Huang 2015). From the two planets orbiting the star P0211 the inner one has a mass of 1.8 \Mjup\ and an orbital period of $P$ = 2.14 day, whereas the outer planet is more massive ($M_{P0211c}\approx$ 7.8 \Mjup), located at a larger
distance from the star ($a_{P0211c}\approx$ 5.5 AU) and in a much more eccentric orbit ($ e \approx$ 0.71) than Pr 0211b (see Table 1). 

There are basically two options for the origin of the structure of this planetary system discussed:

\begin{itemize}
\item Pr 0211 might originally have been surrounded by a planetary system with at least three planets, which has experienced a period of chaotic dynamics leading to planet-planet scattering. Eventually, two planets were left on stable, inner orbits, while the other one was ejected from the system. In the following tidal interaction with the host star circularized the orbit of the inner planet and it became a hot Jupiter at its current position, while the outer planet stayed on an eccentric and misaligned orbit (Weidenschilling \& Marzari 1996;
Chatterjee et al. 2008; Nagasawa et al. 2008).
\item Alternatively, the planetary system experienced a close fly-by of another cluster member. Potentially before existing additional outer planets would have been stripped away and become unbound and the orbit of P0221c was excited to this high eccentricity.
\end{itemize}

Here  we concentrate on the second case. Close stellar fly-bys are expected to be frequent in long-lived clusters, at least in the
initial stages of cluster evolution, possibly influencing the typical orbital architecture
of planets around stars in clusters (Zakamska \& Tremaine 2004; Malmberg \& Davies 2009).
As a consequence, there should be a large number of systems with a hot Jupiter and a second giant planet on an outer eccentric orbit.
 There have been first simulations that try to determine the expected frequency of
such systems (Hao et al. 2013; Li \& Adams 2015) in general.  Shara et al. (2016) even advocate
a cluster origin for several field stars with planetary systems similar to that of Pr 0211.

In contrast to previous work we want to specifically model the situation in M44 and the likelihood of forming a system with the properties found for Pr 0211. In section 2 we describe the numerical method we use to model the cluster dynamics and the effect of fly-bys. In section 3 it is discussed which kind of fly-by would lead to the properties observed for Pr 0211. Then we show how frequent such events are in a cluster like M44.
This is followed by a discussion and summary in sections 4 and 5.

\section{Cluster simulations}

Here  a two-step approach is used to model the effect of the cluster surrounding on discs and/or planetary systems (DPS) similar to our previous work 
\citep{steinhausen:14,vincke:15,vincke:16}. In this approach first the cluster dynamics is simulated and simultaneously the fly-by history recorded. Afterwards, the fly-by history is used 
to determine the effect on the DPS.

\subsection{Method}

 Here we perform cluster simulations representative for M44  using the code Nbody6++GPU (Aarseth 1973; Spurzem 1999;
Aarseth 2003). Our simulations start when the cluster is fully formed and we assume that the cluster stays embedded  in the gas and dust it formed from for another 1 Myr.  This gas is not treated explicitly but just as a background potential.  For the general cluster simulation parameters, see Table 2. As this investigation is only supposed to give an estimate of the
frequency of close fly-bys we did take  into account neither tidal forces nor stellar evolution. In a follow-up study these effects should be included.

The majority of young cluster disperse their stars within the first 10 Myr and its members become part of the field star population (Lada \& Lada 2003, Fall et al. 2009; Dukes \& Krumholz 2012), only the ones that were initially quite compact and sufficiently massive
will survive beyond 10 Myr and only about 4\% survive beyond 100 Myr in the Milky Way (Lada \& Lada 2003).  
Examples of clusters currently younger than 10 Myr that are likely to survive beyond 100 Myr are Westerlund 1 and Trumpler 14
\citep{pfalzner:13,andersen}.

The density in young clusters changes rapidly within the first 20 Myr after their formation.   It can change by several orders of magnitude during that timespan and therefore the present day stellar density of M44 is probably not representative for its value in the past.  Neither is its mass, because one can expect that M44 lost about half of its members over the past 790 Myr. Therefore we assume that the cluster existed initially of about 4000 stars. M44 had its densest phase when it was just a first few Myr old.

In simulations the initial cluster density is characterized by the initial half-mass radius, which becomes the key property when simulating the fly-by frequency in the early phases of cluster development. Often the initial half-mass radius is approximated by one observed in some selected cluster. Unfortunately so far no massive compact cluster has been observed that is (nearly) fully formed but still embedded in its gas. All observed young, massive, compact clusters are basically gas-free so that it can be expected that they had even smaller half-mass radii before gas expulsion \citep{bastian:06}.  However, looking at massive compact clusters in the Milky Way that are younger than 20 Myr one can see that their sizes develop in a well-defined specific way \citep{pfalzner:09}.  The combined knowledge of this size evolution and the mass development allows to determine the initial conditions at the point of gas explusion \citep{pfalzner:13}. These calculations show that the clusters had a sizes in the range 0.1-0.3 pc before they expanded due to gas expulsion and stellar ejections.
In addition, these calculations show that the star formation efficiency (SFE), that is the fraction of gas in the cluster which is turned into stars, for compact clusters is much higher  (60-80\%) than for the clusters in the solar neighbourhood (30\%). This high SFE in compact clusters has been confirmed by observation \citep{rochau:10,cottaar:12}. The high SFE means that although the gas expulsion process 
leads to a portion of its members becoming unbound, the total mass loss is much less dramatic than for the typical clusters in the solar neighbourhood. This is basically the reason for the long-lasting nature of the cluster. Nevertheless, the cluster expands significantly - by a
factor $\approx$ 10 -, however, the main reason is the ejection of stars from the densest cluster regions rather than gas expulsion \citep{pfalzner:13}. Therefore we adopt an initial half-mass radius of 0.2 pc and chose the SFE  to be 70\%.

\begin{table}[t]
% \begin{indented}
\begin{tabular}{@{}lllllll}
\hline
 Parameter & $N_ {stars}$ &  $N_ {sim}$ & SFE & $_ {hm} [pc]$ &$t_ {em}$ [Myr]\\
\hline
                & 4 000           &  50             & 0.7               & 0.2          & 1.0\\
\hline
\end{tabular}
%\end{indented}
 \caption{\label{Table.Two} {Cluster simulation parameters}}
\end{table}

For the mass distribution in the cluster we choose  a modified King profile for the stars and a corresponding Plummer profile for the gas,
because this reflects the situation in observed clusters very well (Espinoza et al. 2009; Steinhausen 2013).   No primordial binaries were included, as this would have significantly
complicated  the determination of the effect of a fly-by on the system. We assumed the simplest initial conditions, that is, the cluster being initially in virial equilibrium, no sub-structure and no mass segregation. The latter two would lead to additional close fly-bys so that the close fly-by frequency found here can be regarded as lower limit \citep{parker:15}. 
For a more detail discussion of the cluster initial conditions see Vincke et al. (2015).

Another simplification made,  is that we did only simulate the first 50 Myr of the dynamical evolution of M44. The reason is that such simulations are computationally expensive.  We found that after 20 Myr the  cluster density changes decreases only slightly (see Fig.1, where temporal development of the half-mass radius is shown ). This means that the dynamical evolution of the cluster is very slow and as a consequence the frequency of close fly-bys basically does not change any more. However, fly-by frequencies during the early phases $<$ 10 Myr depend strongly on the actual set up even for the same parameters but just different seeds. Therefore we opted for performing 11 simulations with different seeds for 50 Myr rather than fewer simulations for a longer timespan. This way we obtain
statistically relevant results for the phase when most close fly-bys happen. We also did not include stellar evolution, since this is of minor importance during the first 50 Myr. In future all these effects should be included, but as we aim only at an estimate of the likelihood of the planetary system around Pr 0211 being shaped by a fly-by, this simplified treatment should suffice. 
Additional details on the numerical method of the cluster simulation including a discussion on the approxiamtions 
can be found in Vincke \& Pfalzner (2017).   

\subsection{Cluster development}

As we will see in section 3, the frequency of close fly-bys is highest during the first 10 Myr and remains fairly constant afterwards. 
The actual frequency can vary considerably from setup to setup \citep{parker:12}. As these type of simulations are generally computationally expensive we opted for
simulating only the first 50 Myr but doing several simulations to obtain statistically significant results for the important early period. Only for one simulation we simulated
the full timespan. As we did not find a significant deviation from our predictions, this method seems adequate.

\begin{figure}[t]
\includegraphics[width=0.45\textwidth]{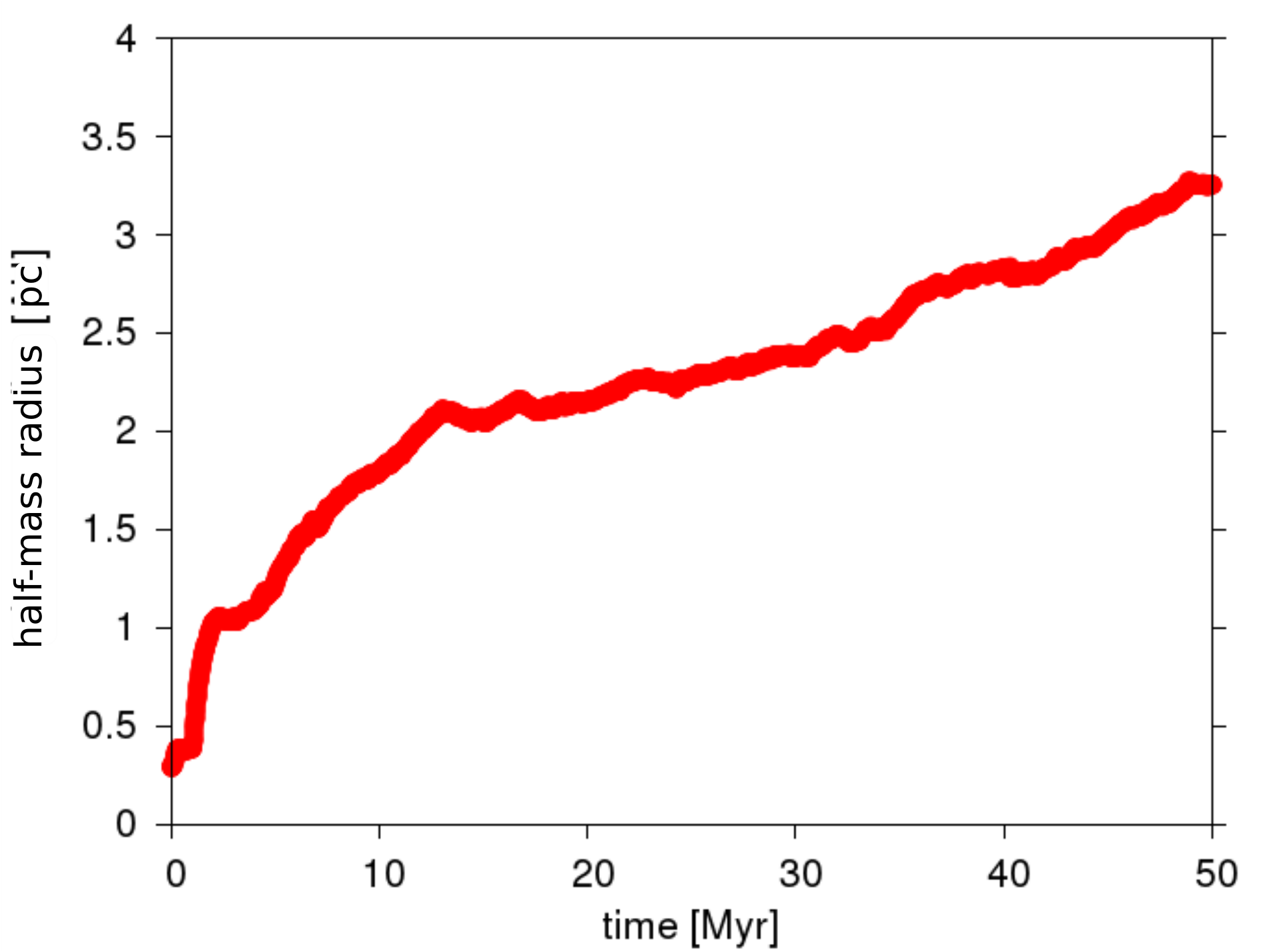}
\caption{Temporal development of the half-mass radius of the simulated M44-like cluster. The presented result is obtain by averaging over 11 different realizations of this cluster.}
\label{fig:cluster_half_mass_radius}
\end{figure}

Fig. 1 shows the development of the cluster half-mass radius over time. The clusters expand from an initial radius of  0.2 pc to about 3 pc over the first 50 Myr.
As most clusters older than 100 Myr in the Milky Way have typically half-mass radii in the range 2-3 pc this seems a realistic representation of the development of M44. 
If anything the cluster expands too little much as including the here not treated binaries and stellar evolution could lead to additional cluster expansion. This
means that again the fly-by frequency at older ages is underestimated rather than overestimated. Thus the following results should be regarded as lower limits
of the occurrence of such close fly-bys.

\begin{figure*}[t]
\includegraphics[width=0.98\textwidth]{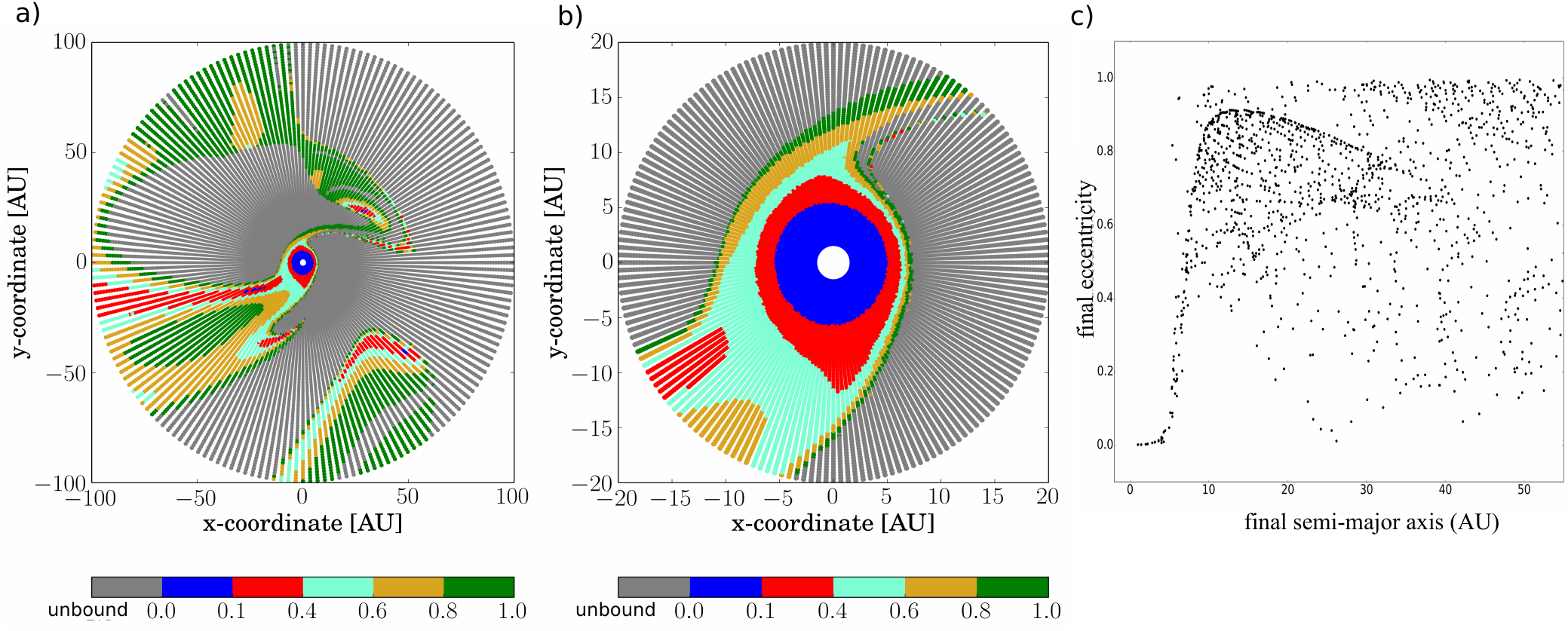}
\caption{The fate of matter in the DPS after a fly-by here shown for the typical case of a 20 AU periastron distance. The stars were of equal mass and the orbit was parabolic and coplanar.   The colour indicates the eccentricity after the fly-by. The position of the matter corresponds to that if the star would have remained unperturbed, but at the the time of periastron passage. a) shows the fate of the particles in the entire disc, b) zooms in onto the relevant area, and c) shows the final eccentricity as a function of the final semi-major axis. }
\label{fig:cluster_half_mass_radius}
\end{figure*}

\section{Effect on the system size and eccentricity}

\subsection{Method}

After the cluster simulations were performed,  the recorded fly-by history was used to determine the effect on the DPS.   Here we applied the results from our previous studies of the effect of fly-bys on discs sizes where we neglected viscous forces and self-gravity within the disc \citep{pfalzner:05}. This means also that we neglect the effect of viscous spreading in the discs. As such, the disc size remains constant throughout our simulations unless altered by a consecutive fly-by (cf. Rosotti et al. 2014). However, neglecting viscosity has the advantage that the same method can be applied in the protoplanetary disc phase as well as the planetary system phase. Another advantage is that existing studies cover a very wide parameter space for fly-by which can be made use of. 

It was assumed that initially each star was surrounded by a protoplanetary disc of 200 AU and an equivalent size was anticipated for the potentially existing planetary system.
This value is motivated by the fact that in Taurus, a prototype of a sparsely populated region, the disc size distribution peaks at 200 AU  \citep{andrews:07}. This should be representative 
for systems that have not been processed by their surroundings.  Similar direct observations of planetary systems with planets on wide orbits mainly show periastra in the range of 100-200 AU  \citep{mcc.96,eisner:08,bally:15}.  In section 3 it will be shown that the majority of discs/planetary systems in M44 were stripped to sizes well below 100 AU, so the result hardly depend on the initial disc-size choice. The likely high stellar density in M44 mean that the disc size might have been altered not only by stellar fly-bys but also by photo-evaporation during the early phases. This has not been taken into account here but should be considered in a follow-up study.

When recording the fly-by history we only take into account events  that lead to a disc-size reduction of at least 5\% ($r_{disc}/r_{previous } \leq $ 0.95), because this is generally the error range of the parameter studies utilized here. Another simplification is that the mass transport from the perturber is neglected. The reason is that captured matter is usually deposited very close to the star and does not influence the disc size as such (Pfalzner et al. 2005).

\subsection{Results}

The general effect of a fly-by on a disc is that some particles remain unaffected, others acquire sub- or super-keplerian velocities and move onto different orbits, some become 
unbound and  a fraction might even become bound to the perturbing star.  Here we concentrate on the matter that remains bound to the host star.  The matter that remains unaffected is mainly close to it, remains on fairly circular orbits and Pr0211b could be an example for a planet in this region.  Matter at larger distances from the
host is more effected by the perturber, and it is often propelled onto highly eccentric orbits. Breslau et al. (2015) have shown that the central unperturbed area mainly determines the disc size and therefore the planetary system size. There is only a narrow transition area where there is still a considerable amount of matter on eccentric orbits whereas outside this
area there is relatively little mass with a few particles moving on highly eccentric orbits. The latter is similar to our Kuiper belt or the transneptunian objects in general. The relative location of these areas depends strongly on the actual fly-by parameters.  

How can we apply this knowledge to determining a possible fly-by scenario for Pr0211? The outer planet Pr0211c would have to belong
to the transition area, between the part of the DPS that remains unperturbed and the part that becomes unbound due to the fly-by. 
It is often assumed that the fate of matter in the DPS after the fly-by is simply a function of its distance to the central star. This is a
crude over-simplification. The actual fate of
the disc material or planet is not only a question of the periastron distance of the perturber passage but also the mass ratio between the two stars involved and, most importantly, sensitively on the relative position at periastron passage. This is illustrated by Fig. 2, which shows the eccentricity of matter after the
fly-by as a function of projected position at periastron distance (for additional information on this type of representation see Breslau et al. 2017).

Pr0211c has a semi-major axis of 5.5$^{+3.0}_ {-1.4}$ AU and moves on a highly eccentric orbit. This means that the inner unperturbed area cannot
exceed $\approx$ 6 AU and is probably even smaller.  It is obvious that only very close fly-bys will lead to system sizes smaller than 6 AU. How close the fly-by had to be to lead to such a small
system size depends on the mass ratio between the star and the perturber and the inclination of the orbit. To
obtain an idea for the envisaged parameter space we first look only at coplanar fly-bys. In this case the following relation  
$ r_d = 0.28 \cdot m_{12}^{-0.32} r_{peri} $ holds \citep{breslau:2014}, where $m_{12}$ is the ratio between mass of the perturber and that of the host star and  $r_{peri}$ the periastron distance of the fly-by. With $r_d$ = 6 AU this leads to a periastron distance
\be  r_{peri} = (20.2) \cdot m_{12}^{0.32}. \ee
If the perturber was of equal mass as Pr0211, it would have had to pass at $\approx$ 20 AU or less to lead to such a small
system.  Fig. 3 gives the region in mass-radius plane where fly-bys would lead to such small system sizes.

\begin{figure}[t]
\includegraphics[width=0.48\textwidth]{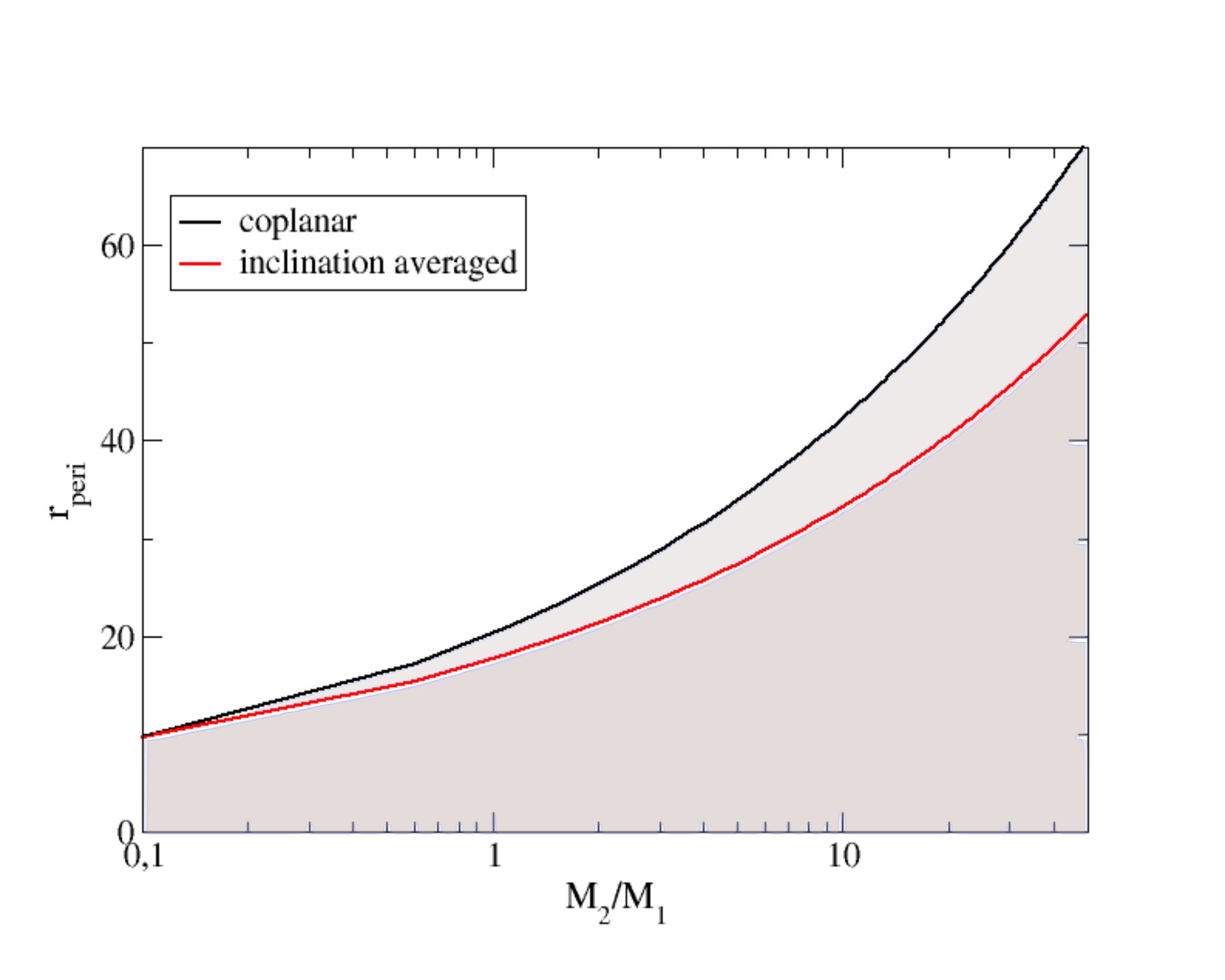}
\caption{Parameter space where stellar fly-bys lead to DPS sizes of 6 AU or less. The black line shows the case of a coplanar event and the red line the case when it is averaged over all possible orientations between the plane of the DPS and the fly-by plane. In both cases a parabolic orbit was assumed. }
\label{fig:6AU}
\end{figure}

It can be expected that in many cases the fly-by would be not coplanar, but move on an inclined orbit. Then the effect on the disc size is less and it would have needed to be even closer to result in $r_d <$ 6 AU. Bhandare et al.(2016) and Vincke \& Pfalzner (2017) give two different approximation for an inclination angle-averaged
disc size after a fly-by. Here we use the formula given by Vincke \& Pfalzner (2017), which
leads to the following formula for the necessary periastron distance
\be  r_{peri} = (1.6 \cdot m_{12}^{-0.2} -1.26  \cdot m_{12}^{-0.182})^{-1} r_{d} \ee
The relevant parameter space is also illustrated in Fig. 3 by the shaded area.

Returning to Fig. 2, this shows the actual fate of the disc matter for a coplanar fly-by of a equal-mass perturbrt  ($m_{12}$=1) that leads to a disc size of 6 AU.   Fig. 2a) shows the fate  of all particles in the DPS.  The colours indicate the eccentricity of the DPS matter after the fly-by, where the blue areas show matter that remains basically unperturbed by the fly-by, whereas matter in the grey areas would become unbound or even captured by the perturber.  All other matter would be bound on eccentric orbits. Fig. 2a)  mainly illustrates that most of the DPS is lost in such a close fly-by and mainly matter very close to the star remains bound. However, in contrast to commonly held view, a small fraction of matter belonging to the outer parts of the disc does remain bound.  In Fig. 2b)  just the central part is shown in magnification.  Here the matter with eccentricities similar to that of Pr0211c are of special interest, which is shown in light brown. For this particular case of a equal mass coplanar encounter, it is a relatively well defined area where such type of eccentricities are induced. For inclined fly-bys and different $m_{12}
$ the location in the original disc can considerably differ.  Fig. 2c) shows the eccentricity vs. the semi-mayor axis of the matter after the fly-by for the co-planar case. It can be seen that some of the matter lies in the parameter space relevant for Pr 0211. For slightly closer fly-bys we would obtain a better match to the properties of Pr0211c, but we are only attempting an estimate of the frequency of such events, so the determined parameters should suffice.

\begin{figure}[t]
\includegraphics[width=0.48\textwidth]{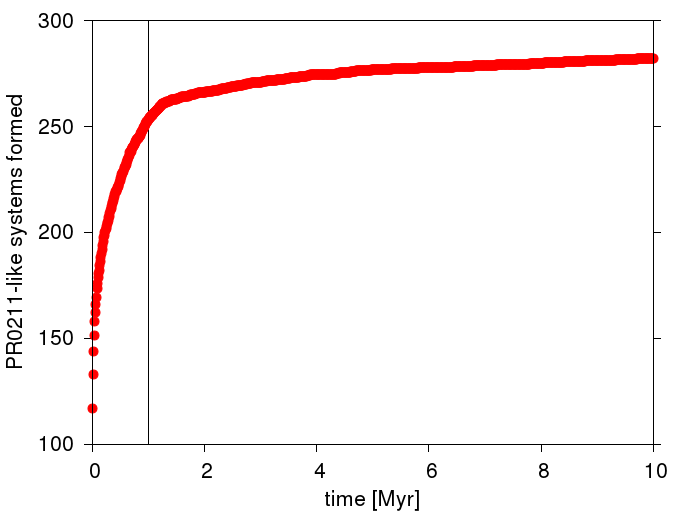}
\caption{Cumulative number of encounters that lead to a system size of less than 6 AU as a function of time since cluster formation.  The horizontal line at 1 Myr indicates the assumed time of gas expulsion.}
\label{fig:rate_age}
\end{figure}

In our cluster simulations of M44 all fly-bys that fulfil the criterion given by Eq. 2 have been tracked. 
Fig. 4 shows the cumulative number of fly-bys that lead to $r_d <$ \mbox{6 AU} as a function of time since cluster formation.
Note that this time is not identical to cluster age as the cluster requires 0.5-2 Myr to form.  Thus to translate the time given here into cluster age one would have to add the formation time.  In Fig. 4 the horizontal line indicates the time of gas expulsion ( here 1 Myr ). It can be seen that the number of fly-bys that lead to such small system sizes is, not surprisingly, highest during the embedded phase. This is simply so, because the stellar density is highest during that phase.  During this phase about 260 systems are cut down to such a small size.  This means that during the embedded phase $\approx$6.5\% of all discs/planetary systems are reduced to sizes of 6 AU or less in a cluster like M44.

\begin{figure}[t]
\includegraphics[width=0.40\textwidth]{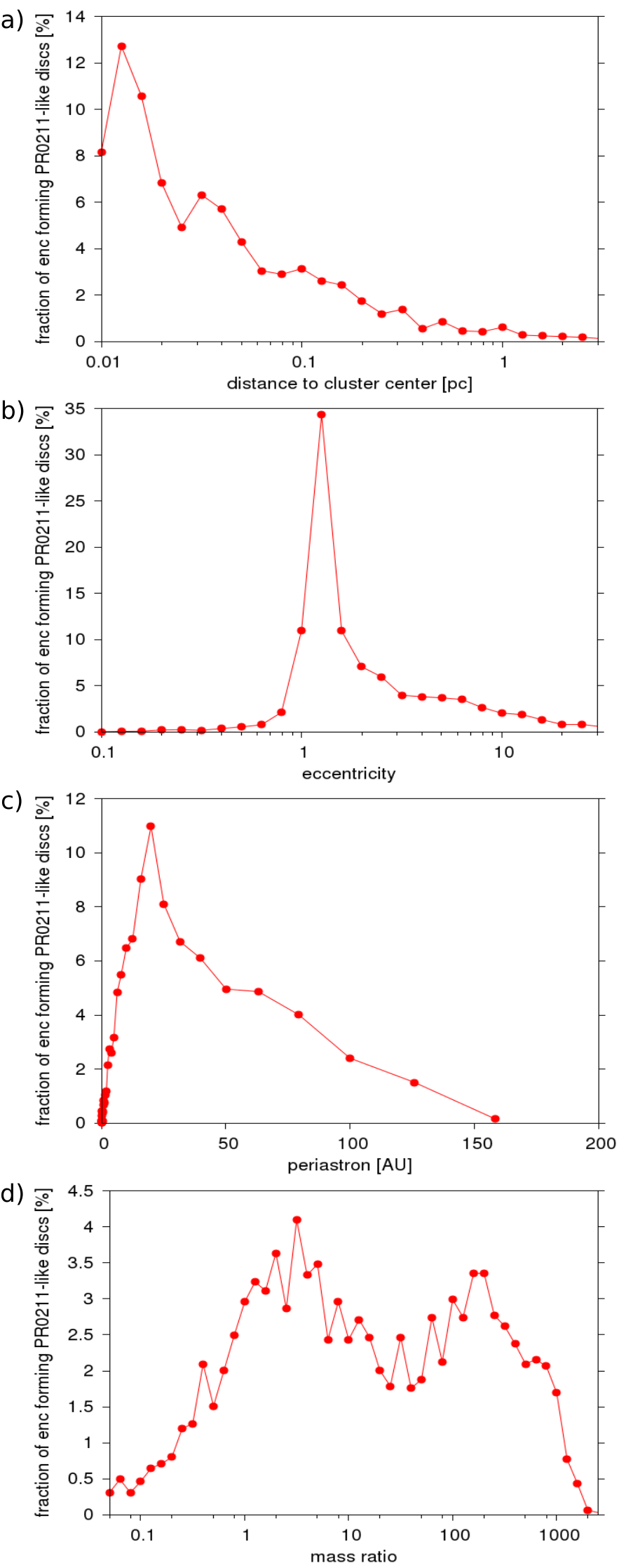}
\caption{Fraction of fly-bys that lead to a system size of less than 6 AU as a function of a) distance to cluster centre, b) eccentricity, c) periastron distance and c) mass ratio between the two stars involved in the fly-by. }.

\label{fig:fly_by_parameters}
\end{figure}

During this embedded phase the cluster expansion is only driven by the loss of stars that are kicked out due to close fly-bys. 
For long-lived clusters gas expulsion is less significant than for the initially more common extended associations, as only $\approx$30\% of the total mass is lost. Therefore, gas expulsion brings the cluster out of equilibrium, but relatively few stars become unbound. Nevertheless, in order to find a new equilibrium state the cluster reacts with considerable expansion, which leads to a drop in stellar density. As a result  the close fly-by rate decreases significantly. Fig. 4 shows that the close fly-by rate after gas expulsion is only about 10\% of its value before gas expulsion.   The cluster finds itself relatively quickly (10 Myr) a new (quasi-)equilibrium state and afterwards the close fly-by rate stays more or less constant at 0.2-0.5 Myr$^{-1}$. If we extrapolate that to the current age of M44, we would expect
12\%-20\% of stars in M44 to have DPS sizes of less than 6 AU due to fly-bys. This value is probably somewhat on the high side, because the cluster will experience stellar evolution and binary interactions, which were not included in this study. However, the fly-by rate will not decrease by more than a factor of two over the following 780 Myr. 

%\begin{figure}[h]
%\includegraphics[width=0.6\textwidth]{cluster_comparison__avg__disc_size_change-200-par-incl-equ-PR0211_like_discs_vs_time.png}
%\caption{Mass of the most massive member of a cluster as function of cluster mass (adapted from Weidner et al. 2010). }
%\label{fig:MW_size_Rgc}
%\end{figure}

Can we say anything about the typical fly-by that leads to a system like that around Pr 0211?
Fig. 5 shows the typical  parameters of close fly-bys in our simulations. Not surprisingly such destructive fly-bys mostly happen close to the cluster centre where the stellar density is highest, more specific the central 0.1 pc of the cluster (see Fig.5a). The vast majority of fly-bys occurs on nearly parabolic orbits (Fig. 5b), which means that our assumption of a parabolic encounter, when calculating the resulting disc size, was justified. This is similar to the situation in more extended clusters/associations typical for the solar neighbourhood \citep{olczak:10}. In contrast, one finds that in more massive likely long-lived clusters like NGC3603, where the stellar denisty is even higher, hyperbolic encounters are more common (Olczak et al. 2012).  We can say actually very little about the likely mass of the perturber star, because the mass-ratio distribution is relatively broad (Fig. 5c). Basically there are two peaks one at around 1-2 and one at high mass ratios, however,
in between the distribution is relatively flat.   By contrast, the periastron range is well-defined with a peak around 20 AU. 

Fig. 6 shows the distribution of disc sizes at 50 Myr of cluster development. It can be seen that the bin with DPS sizes $<$10AU is actually the most populated one.  About 13.5\% of stars have a disc size < 10 AU at 50 Myr. This means that such small system sizes are actually very common.  More then 26\% are smaller than 30 AU the size of our solar system, and virtually no DPS is unaffected by the cluster environment of M44. This means in general that the DPS should be on average much smaller than those of the field stars.

In summary, a fly-by at a distance of $\approx$ 20 AU on a parabolic orbit  when Pr 0211 was close to the cluster centre, would be the most likely scenario.

\section{Application to other clusters and cluster properties}

The question arises to what extend the here presented results for M44 can be generalized to other long-lived clusters. Generally planetary systems in long-lived clusters will be realtively small, which means planets woth $r_P >$ 10 AU will be quite rare. As a result of fly-bys the outer planets of such systems will be often on fairly eccentric orbits. M44 is on the low-mass end of long-lived clusters,
in more massive, and therefore denser,  clusters the influence of close fly-bys will be even more pronounced. Thus one can expect that
the average system size will decrese with increasing cluster mass. Thus if Westerlund 1 really develops in a long-lived cluster its planetary systems can be expected to be even smaller than those in M44.  

Here we discussed the possibility that the orbit of Pr 0221c was caused by the effect of a fly-by on the disc around Pr 0221. 
Another alternative to the here discissed scenario would be that Pr 0211c was capture form the star that flew past. It has been shown that such events are not unlikely for planets/dwarf planets on wider orbit in less dense clusters \citep{lucie:15}. Given the above results the capture scenario would not be far fetch for the Pr 0211 system.   As capture requires approximately the same periastron distance range as the here discussed case, it would require a detailed study which of the two processes is actually the more likely one.

In principle our results could also be used to give an estimate on the number of free floating planets in M44. Pacucci et al. (2013) estimated that 26\% of stars in the cluster could have lost their planets. We take our solar system as a template, meaning with a Jupiter-sized planet at 5 AU, a Uranus-sized planet at 20 AU and a Neptun-sized planet at 30 AU. This might not be the best choice, because exoplanet research has shown that there exist many systems that differ significantly from our own solar-system, but it is the only system where we are sure that we know the full extend of massive planets. Using above results we would expect that M44  contains 160-360 free floating Jupiters, plus 200-600 free-floating Uranuses and an additional 220-800 free-floating Neptunes.

\section{Summary and Conclusion}

The exoplanet system  around  Pr 0211 in the cluster M44 consists of at least two planets with the outer planet moving  on a highly eccentric orbit. Here we tested the hypothesis that a close fly-by of a neighbouring star might be responsible for the eccentric orbit of the second planet by performing simulation of the cluster dynamics. We determined the frequency of fly-bys that would lead to systems equivalent to that around Pr 0211. We find that such close fly-bys are most common during the initial 2-3 Myr after cluster formation. During that time span about 6.5\% of all stars in the cluster would experience a fly-by that would lead to a system size of 6 AU or less.
However, given the uncertainty in the planet formation duration, planets might have not finshed their formation at that time. Thus the fly-by would possibly have needed to take place at later times. However, although the frequency of such fly-bys decreases significantly  due to cluster expansion after gas expulsion, this does {\em not} mean that such a fly-by could not have happened at later times. The reason is that afterwards the fly-by frequency remains  more or less constant at 0.2-0.5 Myr$^{-1}$.  If we extrapolate that to the current age of M44 - 790 Myr - one can expect another 6-14\% of stars to experience such a close fly-by. In other words, it is equally likely that the such a fly-by occured during the first 5 Myr or later on.  Thus in total we would expect 12\%-20\% to have undergone a close fly-by. 
One can conclude that although this is not a definite proof that the high eccentricity of Pr 0211c was caused by a close fly-by, it makes it a convincing option that has to be seriously considered.  

Our simulations show that small system sizes can be expected to be very common in M44. About 14\% of stars should have planetary systems smaller than 10 AU and 27\% systems smaller than our solar system ( 30 AU ) due to stellar fly-bys. These figures are just the lower limit because other effects like external photo-evaporation can lead to an additional reduction of disc sizes during the formation phase. In more massive compact clusters, like for example NGC 3603, the influence of stellar fly-bys would be even stronger, so that there planetary systems like that around Pr 0211 should be very common.

\begin{figure}[t]
\includegraphics[width=0.5\textwidth]{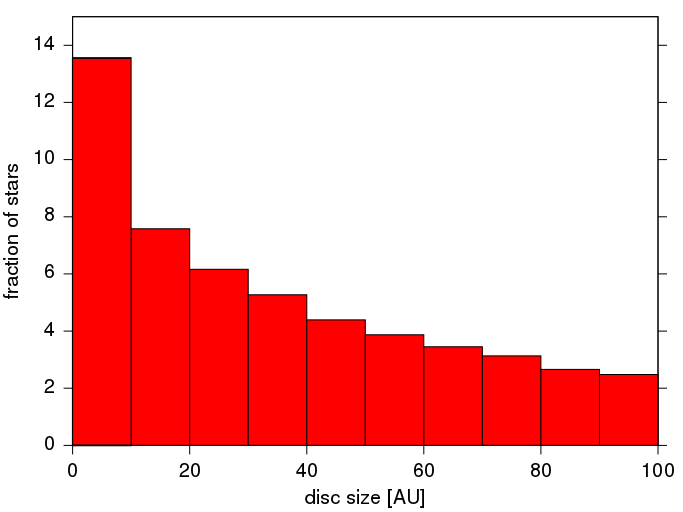}
\caption{Distribution of DPS sizes after 50 Myr of cluster development. }
\label{fig:fly_by_parameters}
\end{figure}

%\begin{figure}[h]
%\includegraphics[width=0.6\textwidth]{cluster_comparison__avg__disc_size_change-200-par-incl-equ-PR0211_like_discs-dsc_enc_vs_ecc.png}
%\caption{Mass of the most massive member of a cluster as function of cluster mass (adapted from Weidner et al. 2010). }
%\label{fig:MW_size_Rgc}
%\end{figure}

%\begin{figure}[h]
%\includegraphics[width=0.6\textwidth]{cluster_comparison__avg__disc_size_change-200-par-incl-equ-PR0211_like_discs-dsc_enc_vs_rel_mpert.png}
%\caption{Mass of the most massive member of a cluster as function of cluster mass (adapted from Weidner et al. 2010). }
%\label{fig:MW_size_Rgc}
%\end{figure}

%\begin{figure}[h]
%\includegraphics[width=0.6\textwidth]{cluster_comparison__avg__disc_size_change-200-par-incl-equ-PR0211_like_discs-dsc_enc_vs_rperi}
%\caption{Mass of the most massive member of a cluster as function of cluster mass (adapted from Weidner et al. 2010). }
%\label{fig:MW_size_Rgc}
%\end{figure}

\bibliographystyle{unsrt}
%\begin{thebibliography}{alpha}

\end{document}